\documentclass[11pt]{article}
\pdfoutput=1

\usepackage{graphicx,amsmath,amssymb,url}

\newcommand{\red}{} 

\newtheorem{theorem}{Theorem}
\newtheorem{definition}{Definition}

\newcommand{\MULTIedge}{    
\begin{picture}(70,30)
\put(5,0){\circle*{5}} \put(5,5){1}
\put(25,0){\circle*{5}} \put(20,5){2}
\put(45,0){\circle*{5}} \put(45,5){3}
\put(65,0){\circle*{5}} \put(65,5){4}
\put(5,0){\line(1,0){20}}
\put(45,0){\line(1,0){20}}
\qbezier(25,0)(35,20)(45,0)
\qbezier(25,0)(35,0)(45,0)
\end{picture}}

\newcommand{\Petersen}{  
\begin{picture}(60,70)
\put(30,60){\circle*{5}}  \put(33,63){1}  \put(0,40){\line(3,2){30}}
\put(60,40){\circle*{5}}  \put(54,45){2}  \put(60,40){\line(-3,2){30}}
\put(50,0) {\circle*{5}}   \put(55,0){3}   \put(50,0){\line(1,4){10}}
\put(10,0) {\circle*{5}}   \put(0,0){4}    \put(10,-1){\line(1,0){40}}
\put(0,40){\circle*{5}}  \put(1,45){5}   \put(10,0){\line(-1,4){10}}
\put(30,49){\circle*{5}}  \put(34,42){6}  \put(30,50){\line(-1,-2){10}}
\put(40,30){\circle*{5}}  \put(44,23){7}  \put(30,50){\line(1,-2){10}}
\put(40,10){\circle*{5}}   \put(44,10){8} \put(40,10){\line(0,1){20}}
\put(20,10){\circle*{5}}   \put(22,1){9}  \put(20,10){\line(1,0){20}}
\put(20,30){\circle*{5}}  \put(9,20){10} \put(20,10){\line(0,1){20}}
\put(30,50){\line(0,1){10}}
\put(40,30){\line(5,3){20}}
\put(40,10){\line(1,-1){10}}
\put(20,10){\line(-1,-1){10}}
\put(20,30){\line(-5,3){20}}
\end{picture}}

\newcommand{\Petersentwo}{ 
\begin{picture}(60,70)
\put(30,60){\circle*{5}}  \put(33,63){1}  \put(0,40){\line(3,2){30}}
\put(60,40){\circle*{5}}  \put(55,45){2}  \put(60,40){\line(-3,2){30}}
\put(50,0) {\circle*{5}}   \put(55,0){3}   \put(50,0){\line(1,4){10}}
\put(10,0) {\circle*{5}}   \put(0,0){4}    \put(10,-1){\line(1,0){40}}
\put(0,40) {\circle*{5}}  \put(0,45){5}   \put(10,0){\line(-1,4){10}}
\put(30,49){\circle*{5}}  \put(34,42){6}  
\put(41,31){\circle*{5}}  \put(44,23){7}  \put(40,30){\line(5,3){20}}
\put(40,10){\circle*{5}}   \put(44,10){8} 
\put(20,10){\circle*{5}}   \put(22,1){9}  
\put(19,31){\circle*{5}}  \put(10,20){10} 
\put(30,50){\line(-1,-4){10}}  \put(30,50){\line(1,-4){10}}
\put(20,30){\line(1,-1){20}}  \put(20,30){\line(1,0){20}}
\put(20,10){\line(1,1){20}}  
\put(30,50){\line(0,1){10}}
\put(40,30){\line(5,3){20}}
\put(40,10){\line(1,-1){10}}
\put(20,10){\line(-1,-1){10}}
\put(20,30){\line(-5,3){20}}
\end{picture}}

\newcommand{\PENTmin}{
\begin{picture}(50,40)
\put(10,20){\circle*{5}} \put(0,20){1}
\put(14,0){\circle*{5}}  \put(5,-5){3}
\put(36,0){\circle*{5}}  \put(40,-5){4}
\put(10,20){\line(1,1){15}}
\put(25,35){\red \line(1,-1){15}}
\put(10,20){\line(1,-5){4}}
\put(40,20){\line(-1,-5){4}}
\put(14,0){\line(1,0){22}}
\put(10,20){\line(6,-5){25}}
\put(25,35){\circle*{5}} \put(30,35){5}
\put(40,20){\red \circle*{5}} \put(45,20){2}
\end{picture}}

\newcommand{\PENTmindel}{
\begin{picture}(50,40)
\put(10,20){\circle*{5}} \put(0,20){1}
\put(14,0){\circle*{5}}  \put(5,-5){3}
\put(36,0){\circle*{5}}  \put(40,-5){4}
\put(25,35){\circle*{5}} \put(30,35){5}
\put(10,20){\line(1,1){15}}
\put(10,20){\line(1,-5){4}}
\put(40,20){\red \line(-1,-5){4}}
\put(14,0){\line(1,0){22}}
\put(10,20){\line(6,-5){25}}
\put(40,20){\red \circle*{5}} \put(45,20){2}
\end{picture}}

\newcommand{\PENTmincon}{
\begin{picture}(50,40)
\put(10,20){\red \line(1,0){30}}
\put(10,20){\circle*{5}} \put(0,20){1}
\put(14,0){\circle*{5}}  \put(5,-5){3}
\put(36,0){\circle*{5}}  \put(40,-5){4}
\put(10,20){\line(1,-5){4}}
\put(40,20){\line(-1,-5){4}}
\put(14,0){\line(1,0){22}}
\put(10,20){\line(6,-5){25}}
\put(40,20){\red \circle*{5}} \put(45,20){2}
\end{picture}}

\newcommand{\PENTV}{
\begin{picture}(50,40)
\put(40,20){\circle*{5}} \put(45,20){2}
\put(25,35){\circle*{5}} \put(30,35){5}
\put(36,0){\circle*{5}}  \put(40,-5){4}
\put(10,20){\line(1,1){15}}
\put(25,35){\line(1,-1){15}}
\put(40,20){\line(-1,-5){4}}
\put(14,0){\line(1,0){22}}
\put(10,20){\line(6,-5){25}}
\put(10,20){\red \circle*{5}} \put(0,20){1}
\put(14,0){\red \circle*{5}}  \put(5,-5){3}
\put(10,20){\red \line(1,-5){4}}
\end{picture}}

\newcommand{\PENTVdel}{
\begin{picture}(50,40)
\put(40,20){\circle*{5}} \put(45,20){2}
\put(25,35){\circle*{5}} \put(30,35){5}
\put(36,0){\circle*{5}}  \put(40,-5){4}
\put(14,0){\circle*{5}}  \put(5,-5){3}
\put(10,20){\line(1,1){15}}
\put(25,35){\line(1,-1){15}}
\put(40,20){\line(-1,-5){4}}
\put(14,0){\line(1,0){22}}
\put(10,20){\red \line(6,-5){25}}
\put(10,20){\red \circle*{5}} \put(0,20){1}
\put(36,0){\red \circle*{5}}  \put(40,-5){4}
\end{picture}}

\newcommand{\PENTVcon}{
\begin{picture}(50,40)
\put(40,20){\circle*{5}} \put(45,20){2}
\put(25,35){\circle*{5}} \put(30,35){5}
\put(36,0){\circle*{5}}  \put(40,-5){4}
\put(10,20){\line(1,1){15}}
\put(25,35){\line(1,-1){15}}
\put(40,20){\line(-1,-5){4}}
\put(10,20){\red \line(6,-5){25}}
\qbezier(10,20)(15,0)(36,0)
\put(10,20){\red \circle*{5}} \put(0,20){1}
\put(36,0){\red \circle*{5}}  \put(40,-5){4}
\end{picture}}

\newcommand{\PENTVmikecon}{
\begin{picture}(50,40)
\put(40,20){\circle*{5}} \put(45,20){2}
\put(25,35){\circle*{5}} \put(30,35){5}
\put(14,0){\circle*{5}}  \put(5,-5){3}
\put(14,0){\line(1,3){11}}
\put(25,35){\line(1,-1){15}}
\put(14,0){\line(1,0){22}}
\qbezier(14,0)(25,10)(36,0)
\put(36,0){\red \circle*{5}}  \put(40,-5){4}
\put(40,20){\red \circle*{5}} \put(45,20){2}
\put(40,20){\red \line(-1,-5){4}}
\end{picture}}

\newcommand{\edge}{
\begin{picture}(30,10)
\put(6,3){\circle*{4}}
\put(24,3){\circle*{4}}
\put(6,3){\line(1,0){18}}
\end{picture}}

\newcommand{\TRIe}{
\begin{picture}(50,40)
\put(10,0){\circle*{6}}
\put(40,0){\circle*{6}}
\put(25,30){\circle*{6}}
\put(35,15){\it e}
\put(10,0){\line(1,0){30}}
\put(10,0){\line(1,2){15}}
\put(25,30){\line(1,-2){15}}
\end{picture}}

\newcommand{\TRIdel}{
\begin{picture}(50,40)
\put(10,0){\circle*{6}}
\put(40,0){\circle*{6}}
\put(25,30){\circle*{6}}

\put(10,0){\line(1,0){30}}
\put(10,0){\line(1,2){15}} 
\end{picture}}

\newcommand{\TRIconbez}{
\begin{picture}(50,30)
\put(10,0){\circle*{6}}
\put(40,0){\circle*{6}}
\put(10,0){\line(1,0){30}} 
\qbezier(10,0)(25,15)(40,0)
\end{picture}}

\begin{document}

\title{A new edge selection heuristic for computing the Tutte polynomial of an undirected graph.}
\author{Michael Monagan\thanks{This work was supported by the Mprime NCE of Canada} \\
Department of Mathematics, Simon Fraser University \\ {\tt mmonagan@cecms.sfu.ca}}

\maketitle
\begin{abstract}
We present a new edge selection heuristic and vertex ordering heuristic that
together enable one to compute the Tutte polynomial of much larger sparse graphs than
was previously doable.  As a specific example, we are able to compute the
Tutte polynomial of the truncated icosahedron graph using our Maple
implementation in under 4 minutes on a single CPU.
This compares with a recent result of Haggard, Pearce and Royle whose special
purpose C++ software took one week on 150 computers.
\end{abstract}


\section{Introduction}

\begin{center}
\begin{figure}[htb]
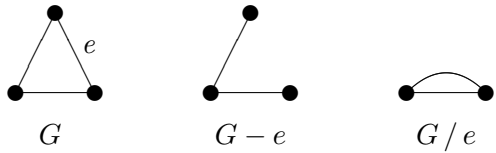
 \centering
\begin{tabular}{c c c c c } \vspace*{2mm}
\TRIe      &&    \TRIdel     && \TRIconbez \\ \vspace*{-2mm}
     $G$   &&      $G-e$     && $G\,/\,e$
\end{tabular} \caption{Graph edge deletion and contraction.} \label{fig:delcon}
\end{figure}
\end{center}

Let $G$ be an undirected graph.
The Tutte polynomial of $G$ is a bivariate polynomial $T(G,x,y)$ which
contains information about how $G$ is connected.  
We recall Tutte's original definition for $T(G,x,y)$.
Let $e=(u,v)$ be an edge in $G$.  Let $G-e$ denote the graph obtained by deleting $e$
and let $G\,/\,e$ denote the graph obtained by contracting $e$, that is, first
deleting $e$ then joining vertexes $u$ and $v$.
Figure \ref{fig:delcon} shows an example of edge deletion and contraction.

\begin{definition}
Let $G$ be a connected undirected graph.  
The {\em Tutte polynomial} $T(G,{\red x,y})$ is the
bivariate polynomial defined by

\begin{equation}
T(G) ~=~ 
  \begin{cases}
   1			& \text{ if $|E|=0$, } \\
   x \, T(G/e)		& \text{ if $e$ is a cut-edge in $G$, } \\
   y \, T(G-e)		& \text{ if $e$ is a loop in $G$ } \\
   T(G-e) + T(G/e)	& \text{ if the edge $e$ is neither a loop nor a cut-edge in $G$. }
  \end{cases}
\end{equation}

\end{definition}

\noindent
This definition immediately gives a recursive algorithm for computing $T(G,x,y)$.
In general, a naive implementation of the algorithm will make an
exponential number of recursive calls because of the last case in (1).
If $G$ has $n$ vertexes and $m$ edges, the number of recursive calls $C(n+m)$ is bounded by
$$C(n+m) \le C(n+m-1) + C(n-1+m-1).$$
This is the Fibonacci recurrence.  Hence $C(n+m) \in O( 1.618^{n+m} ).$
If $G$ is not biconnected one can apply the following theorem to reduce $C(n+m)$.

\begin{theorem} [Tutte \cite{tutte}]
Let $G$ be a graph with $m$ biconnected components (blocks) $B_1, B_2, \ldots, B_m$.  \\
Then $T(G,x,y) = \Pi_{i=1}^m T(B_i,x,y).$
\end{theorem}

Another way to reduce $C(n+m)$ is to ``remember'' the Tutte polynomials computed in the
computation tree and use a graph isomorphism test to test whether a graph in the
computation tree has been seen before.
In \cite{HPR2011}, Haggard, Pearce and Royle present timings for random cubic
and quartic graphs, complete graphs, and random graphs with varying edge densities $0<p<1$
that shows that employing graph isomorphism is very effective.
For example, it roughly increases by 50\% the size of random cubic graphs
that can be handled in a given time.
A factor determining the effectiveness of the isomorphism test is the order in which
the edges are selected.  In \cite{HPR2010}, Haggard, Pearce and Royle investigate various
edge ordering heuristics.  Two heuristics, which they call MINDEG and VORDER,
are found to perform consistently better than random selection.

Our paper is organized as follows.
In section 2 we describe the MINDEG and VORDER heuristics and present a new
edge selection heuristic.  The VORDER heuristic, and our new heuristic, also depend on the
ordering of the vertexes in $G$.  We present an ordering that we have found 
works particularly well with our edge selection heuristic.
In section 3 we describe our Maple implementation and explain how
we test for isomorphic graphs in the computation tree.
In section 4 we present benchmarks comparing the three heuristics with and without
the new vertex ordering and with and without an explicit graph isomorphism test.
The data presented shows that our new heuristic again, roughly increases by 50\% the
size of sparse cubic graphs that can be handled in a given time.
An experimental finding in this paper is that our new edge selection heuristic,
when combined with our vertex ordering, does not require an explicit isomorphism test;
a simple test for identical graphs is sufficient.

We end the introduction with some further information about available software for
computing Tutte polynomials and related polynomials.
Useful references include the very good Wikipedia webpage  \\
http://en.wikipedia.org/wiki/Tutte\_polynomial and Bollob\'{a}s' text \cite{text}.
The graph theory packages in Mathematica and Maple include commands for computing
Tutte polynomials.  The Mathematica algorithm does not look for identical or
isomorphic graphs in the computation tree (see \cite{HPR2011}).
The TuttePolynomial command in Maple 11 and more recent versions (see \cite{farr})
uses the VORDER heuristic and hashing to test for identical graphs in the computation tree.
The fastest available software for computing Tutte polynomials and related polynomials
is that of Haggard, Pearce and Royle \cite{HPR2010,HPR2011}.  
It is available on David Pearce's
website at \verb+http://homepages.ecs.vuw.ac.nz/~djp/tutte/+.
It uses the canonical graph ordering available in Brendan Mckay's nauty package
(see \cite{mckay}) to identify isomorphic graphs.

We recall the definition for the reliability polynomial and chromatic polynomial.
\begin{definition}
Let $G$ be an undirected graph.  The {\em reliability polynomial} of $G$,
denoted $R_p(G)$, is the probability that $G$ remains connected
when each edge in $G$ fails with probability $p$.
The {\em chromatic polynomial} of $G$, denoted $P_{\lambda}(G)$,
counts the number of ways the vertexes of $G$ can be colored with $\lambda$ colors.  
\end{definition}

For example, $R_p(\edge) = 1-p$ and $P_{\lambda}(\edge) = \lambda (\lambda-1).$
The reliability and chromatic polynomials can also be computed by the edge deletion and
contraction algorithm (see \cite{royle}). 
If $G$ has $n$ vertexes and $m$ edges, they are related
to the Tutte polynomial as follows:

\begin{equation}
R_p(G) = (1-p)^{(n-1)} \, p^{(m-n+1)} ~ T(G,1,p^{-1}),
\end{equation}

\vspace*{-3mm}

\begin{equation}
P_{\lambda}(G) = (-1)^{(n-1)} \, \lambda ~ T(G,1-\lambda,0).
\end{equation}

\noindent
Since graph coloring is NP$-$complete, it follows that computing the
the chromatic polynomial is NP$-$hard.  Thus (3) implies computing
the Tutte polynomial is also NP$-$hard.  It is also known that
computing $R_p(G)$ is NP-hard (see \cite{Oxley}).
This does not mean, however, that computing the Tutte polynomial for a given graph
is not polynomial time.  Our new edge selection heuristic is polynomial
time for some structured sparse graphs.

\section{Edge selection heuristics.}

In applying the identity $T( G ) = T(G-e) + T(G/e)$
we are free to choose any edge which is neither a cut-edge nor a loop.
[If $G$ has a cut-edge or loop, then those edges should be processed first.]
In \cite{HPR2010}, Haggard, Pearce and Royle propose two heuristics,
the minimum degree heuristic (MINDEG) and the vertex order heuristic (VORDER).
We describe the heuristics here and introduce our new heuristic which
is a variation on VORDER.

\subsection{The minimum degree heuristic: MINDEG}
\begin{center}
\begin{figure}[htb]
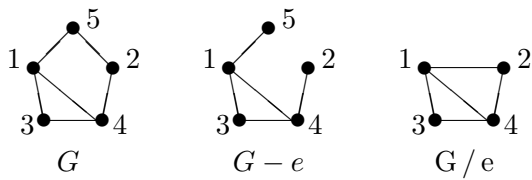
 \centering
\begin{tabular}{c c c c c}
\vspace*{2mm}
\PENTmin   &&   \PENTmindel  && \PENTmincon \\ \vspace*{-2mm}
     $G$   &&      $G-e$     && G\,/\,e
\end{tabular} \caption{The minimum degree heuristic.} \label{fig:mindeg}
\end{figure}
\end{center}

Consider the graph $G$ in Figure \ref{fig:mindeg}.

\vspace*{-2mm}

The minimum degree heuristic picks the edge $e=(u,v)$ where $u$ is the first vertex
of minimum degree ($u=2$ in the example) and $v$ is the first vertex adjacent to $u$
of minimum degree ($v=5$ in the example).  Hence $e=(2,5)$ is chosen.
Shown in the figure are the graphs $G-e$ and $G\,/\,e$.  The reader can see that
the next edge that will be selected in $G-e$ is the edge $(2,4)$, which is a cut-edge.
The algorithm will then contract the edge $(2,4)$, then select the edge $(1,5)$,
another cut-edge.  After contracting $(1,5)$ what is left is the triangle
on vertexes $1,3,4$.  For the graph $G\,/\,e$, the MINDEG heuristic selects the edge
(2,1).  After deleting (2,1), MINDEG will select and contract the edge (2,4) again yielding
the triangle $1,3,4$.  This example shows how identical graphs in the computation
tree arise.

\subsection{The vertex order heuristic: VORDER}
\begin{center}
\begin{figure}[htb]
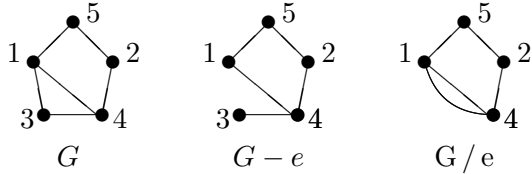
 \centering
\begin{tabular}{c c c c c}
\vspace*{2mm}
\PENTV     &&   \PENTVdel    && \PENTVcon \\ \vspace*{-2mm}
     $G$   &&      $G-e$     && G\,/\,e
\end{tabular} \caption{The VORDER-pull heuristic.} \label{fig:vorderpull}
\end{figure}
\end{center}

Consider again the graph $G$ shown in the Figure \ref{fig:vorderpull}.
The vertex order heuristic picks the edge $e=(u,v)$ where $u$ is simply
the first vertex in the $G$ and $v$ is the first vertex adjacent to $u$.
In our example $u=1$, $v=3$, hence $e=(1,3)$ is chosen.
Shown in Figure \ref{fig:vorderpull} are the graphs $G-e$ and $G\,/\,e$ where
when we contracted the edge $e=(1,3)$ we ``pulled'' vertex 3 down to vertex 1.
The next edge selected in $G \,/\, e$ will be one of the edges (1,4).

\vspace*{-2mm}

\begin{center}
\begin{figure}[htb]
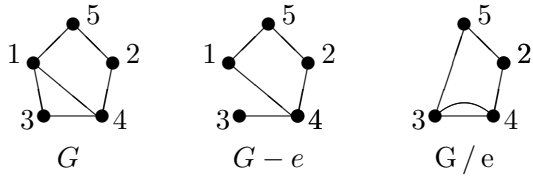
 \centering
\begin{tabular}{c c c c c c}
\vspace*{2mm}
\PENTV     &&   \PENTVdel    && \PENTVmikecon \\ \vspace*{-2mm}
     $G$   &&      $G-e$     && G\,/\,e
\end{tabular} \caption{The VORDER-push heuristic.} \label{fig:vorderpush}
\end{figure}
\end{center}

There is alternative choice here when constructing the graph $G \,/\, e$.
Instead of ``pulling'' vertex $v=3$ down to $u=1$, if instead we ``push'' vertex $u=1$ up to $v=3$
we get the contracted graph shown in Figure \ref{fig:vorderpush}.
Observe that the two contracted graphs $G \,/\, e$ in
figures \ref{fig:vorderpull} and \ref{fig:vorderpush} are isomorphic.
However, in the vertex order heuristic, the next edge
selected in $G\,/\,e$ is different.  In figure \ref{fig:vorderpull} the vertex order heuristic
selects edge (1,4).  In figure \ref{fig:vorderpush} it selects edge (2,4).
We will call the two vertex order heuristics VORDER-pull and VORDER-push, respectively.

\vspace*{-2mm}

To visualize the difference between VORDER-pull and VORDER-push, picture the computation tree 
of graphs produced by the algorithm as it applies the identity $ T(G) = T(G-e) + T(G\,/\,e). $
On the left of the computation tree we repeatedly delete edges.  On the right of the tree we
repeatedly contract edges.  The two vertex order heuristics differ when we contract.
In the VORDER-pull heuristic, we always select the same first vertex
and contract (pull) other vertexes to it thus typically increasing the degree of the first vertex.
In the VORDER-push heuristic, we select the first vertex and push it away (into the middle
of the graph) and move on to the next vertex in the ordering.  Thus one measurable 
difference between VORDER-pull and VORDER-push is that the degree of the
vertex $u$ selected will generally be greater in VORDER-pull than in VORDER-push.
We will measure this explicitly in our benchmarks.

\subsection{The vertex label ordering}

The VORDER-pull and VORDER-push heuristics, and also to a lesser extent,
the MINDEG heuristic, also depend on the input permutation of the labels of the vertexes in $G$.
All three heuristics are sensitive to this ordering with a random ordering producing
a bad behavior.  In \cite{HPR2010}, Haggard, Pearce and Royle state ``using an
ordering where vertexes with higher degree come lower in the ordering generally 
also gives better performance''.  Their idea is to increase the probability that
more identical graphs appear {\em higher} in the computation tree.
To achieve this we propose to label the vertexes in the input graph in an order so that the 
algorithm deletes and contracts edges {\em locally}.
We found that the following vertex ordering heuristic works best amongst the orderings we tried.
To simplify the presentation we assume $G$ is connected.
We describe it below with pseudo-code and an example.

\begin{quote}
Algorithm {\bf SHARC} - short arc order. \\
Input: An undirected connected graph $G$ on $n>0$ vertexes $V = \{1,2,...,n\}$. \\
\begin{itemize}
\item[1] Initialize the ordered list $S = [1]$
\item[2] {\bf while} $|S| < n$ do the following
\begin{itemize}
\item[ ] Using breadth first search (BFS), starting from the vertexes in $S$ find
         the first path from $S$ back to $S$ which includes at least one new vertex,
         that is, find a path $u \rightarrow v_1 \rightarrow v_2 \rightarrow ... \rightarrow v_m \rightarrow w$
         where $u\in S$, $w \in S$, $m>0$, $v_i \in V\backslash S$.

         If such a path exists, append $v_1,v_2,...,v_m$ to $S$.
         Otherwise ($G$ may have a cutedge) pick the least vertex $v_1$
         not in $S$ but adjacent to a vertex in $S$ and append $v_1$ to $S$.
\end{itemize}
{\bf end while}
\item[3] {\bf output} $S$.
\end{itemize}
\end{quote}

\noindent
We explain the algorithm with an example.  Consider again the graph $G$ below.

\begin{center} \PENTV \end{center}

\noindent
Initially we have $S=[1]$.
Using BFS we insert all vertexes adjacent to the vertexes in $S$ not already in $S$ 
into a queue $Q$.  In the example, we obtain $Q = [3,4,5]$.
Hence we have paths $1\rightarrow 3$, $1\rightarrow 4$ and $1\rightarrow 5$ which
we maintain in an array $P=[0,0,1,1,1]$, that is $P_3=1$ stores the edge from 3 to 1
and $P_1=0$ indicates the end of a path.
We take the first vertex $3$ from $Q$ and consider the new edge $(3,4)$.
Since $P_4$ is not zero we know there is a path from $1$ back to $4$ stored in $P$.
Since $3$ came from $Q$ we know there is a path from $1$ to $3$ stored in $P$.
Thus we are done this iteration; we extract the path $1\rightarrow 3 \rightarrow 4 \rightarrow 1$
from $P$ and append $3,4$ to $S$ obtaining $S=[1,3,4]$.
In the second iteration the algorithm will find the
path $1 \rightarrow 5 \rightarrow 2 \rightarrow 4$ and set
$S = [1,3,4,5,2]$.  Since $|S|=5$ the algorithm stops.
The reader can see that the algorithm finds a short cycle in
the first iteration, then in the subsequent iterations, finds short arcs from $S$ back to $S$.
We will call this ordering a short arc ordering (SHARC).
By picking the first path found using BFS, the short arc ordering
maintains locality in $S$.  Although it would be simpler to order the vertexes in simple breadth
first search order, that ordering did not prove to be as good as SHARC in our experiments.

\section{Maple Implementation}

\begin{center}
\begin{figure}[htb]
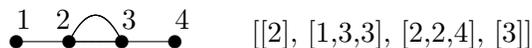
 \centering
\begin{tabular}{c c c }
\MULTIedge  & & [[2], [1,3,3], [2,2,4], [3]]
\end{tabular} \caption{Maple list of lists data structure for $G$} \label{fig:maple}
\end{figure}
\end{center}

We use a list of neighbors representation for a multi-graph in our Maple
implementation.  We illustrate with an example in Figure \ref{fig:maple}.

\noindent
To identify identical graphs in the computation tree we
make use of \verb+option remember+.  This is a feature of the
Maple programming language that enables our Maple procedure
to automatically identify identical graphs in the computation tree using hashing.
For this to work we must canonically re-label vertexes to be $1,2,...,n-1$ after
edge contraction.

To identify non-equal isomorphic graphs we have implemented our own graph isomorphism
test for multi-graphs as the \verb+IsIsomorphic+ command in Maple's \verb+GraphTheory+ package
(see \cite{farr}) treats simple graphs only.
Instead of searching all previous graphs, we first hash on the characteristic polynomial
of the Laplacian matrix of $G$, a known graph invariant.  The Laplacian matrix is an $n$ by $n$
matrix $D-A$ where $D$ is the degree matrix of $G$ and $A$ is the adjacency matrix of $G$.
For increased efficiency, we compute the characteristic polynomial of $D-A$
modulo a machine prime $p$.  This can be computed in $O(n^3)$ arithmetic operations
in $\mathbb{F}_p$.  See Algorithm 2.2.9 in Chapter 2 of \cite{Cohen}.

\medskip
Our Maple code may be downloaded from \verb+http://www.cecm.sfu.ca/~mmonagan/tutte+

\section{Experiments}
\subsection{Random cubic graphs}

In this experiment we generated ten random connected cubic graphs on $n$
vertexes for $16 \le n \le 50$.
Note, the probability that these graphs are biconnected is high so Theorem 1 is not applicable.
Indeed all the graphs generated are biconnected.
We computed the average and median time it takes our Maple program to compute the Tutte polynomial
using the MINDEG, VORDER-pull and VORDER-push heuristics, on a 2.66 Ghz Intel Core i7 980 desktop with 6 GB RAM.
We do this for two permutations of the vertex labels, random (see Table \ref{tab:random}) and SHARC 
(the short arc ordering) (see Table \ref{tab:SHARC}).  In all cases, we do not use an explicit graph isomorphism
test; rather, we use Maple's {\tt option remember;} facility so that Tutte polynomials for identical graphs
that appear in the computation tree are not recomputed.
The data shows the SHARC ordering is much better than the input random ordering for both VORDER-pull and VORDER-push.
The data also shows that VORDER-push with SHARC is much better than VORDER-pull with SHARC.
We find similar results for random quartic graphs.

\begin{table}[htb]
\begin{center}
\begin{tabular}{|l l | r r | r r|  r r |}
\hline
&&  \multicolumn{2}{|c|}{MINDEG heuristic} & \multicolumn{2}{|c|}{VORDER pull} & \multicolumn{2}{|c|}{VORDER push}  \\ \hline
$n$&$m$&  ave  &  med    &  ave   &  med   &   ave  &    med \\ \hline
16& 24&   0.41 &  0.36   &   0.18 &   0.11 &   0.22 &   0.14 \\
18& 27&   1.21 &  1.02   &   0.53 &   0.33 &   0.57 &   0.45 \\   
20& 30&   3.90 &  3.38   &   1.27 &   1.02 &   1.86 &   1.46 \\
22& 33&  14.40 & 12.07   &   4.65 &   3.36 &   7.22 &   6.88 \\
24& 36&  56.24 & 32.19   &  13.84 &   9.23 &  25.05 &  22.46 \\
26& 39& 193.34 & 118.98  &  41.03 &  20.07 &  58.94 &  24.57 \\
28& 39&        &         & 199.70 & 116.32 & 210.69 &  75.24 \\ \hline
\end{tabular}
\end{center}
\caption{Timings in CPU seconds for random cubic graphs with $n$ vertices using random vertex order.} \label{tab:random}
\end{table}

\begin{table}[htb]
\begin{center}
\begin{tabular}{|l l | r r | r r|  r r |}
\hline
  &   &  \multicolumn{2}{|c|}{ MINDEG heuristic }  & \multicolumn{2}{|c|}{VORDER pull}  &  \multicolumn{2}{|c|}{  VORDER push }  \\ \hline
$n$&$m$&   ave  &  med    &  ave   &   med  &   ave  & med   \\ \hline

 18&  27&   0.68 &  0.51   &  0.05  &  0.03  & 0.02   & 0.02  \\
 22&  33&   7.73 &  4.68   &  0.38  &  0.14  & 0.10   & 0.07  \\
 26&  39&  80.11 & 38.45   &  1.24  &  0.41  & 0.17   & 0.12  \\
 30&  45&        &         & 11.10  &  4.36  & 0.67   & 0.37  \\
 34&  51&        &         & 94.58  & 19.15  & 2.06   & 1.29  \\
 38&  57&        &         &        &        & 5.40   & 2.83  \\
 42&  63&        &         &        &        & 40.66  & 8.82  \\
 46&  69&        &         &        &        & 87.63  & 49.03 \\
 50&  75&        &         &        &        & 179.64 & 39.61 \\  \hline
\end{tabular}
\end{center}
\caption{Timings in CPU seconds for random cubic graphs with $n$ vertices using SHARC vertex order.} \label{tab:SHARC}
\end{table}

\subsection{Generalized Petersen graphs.}

\begin{figure}[htb]
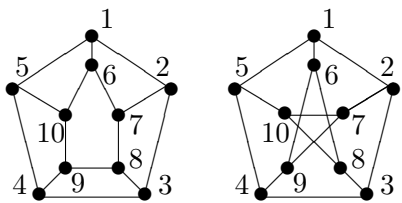
 \centering
\begin{tabular}{c c c}
\Petersen & & \Petersentwo
\end{tabular} \caption{Petersen graphs $P(5,1)$ and $P(5,2)$.} \label{fig:peter}
\end{figure}

The generalized Petersen graph $P(n,k)$ with $1 \le k < n/2$ is a cubic graph on $2n$
vertexes.  Figure \ref{fig:peter} shows $P(5,1)$ and $P(5,2)$.  $P(5,2)$ is the familiar Petersen graph.
To construct $P(n,k)$ the vertexes are divided into two sets $1,2,...,n$ and $n+1,n+2,...,2n$,
which are placed on two concentric circles as shown in figure \ref{fig:peter}.
The first set of vertexes are connected in a cycle $1,2,...,n,1$.
The second set are connected to the first with vertex $i$ connected to $n+i$ for $1 \le i \le n$.
The second parameter governs how the second set is connected.
Connect $n+i$ to $n + (n+i \pm k \mod n)$ for $1 \le i \le n$.

The SHARC vertex order for $P(5,1)$ and $P(5,2)$ is [1,2,7,6,10,5,4,3,8,9]
and [1,5,4,3,2,8,6,9,10,7] respectively.
In Tables \ref{tab:P3pull} and \ref{tab:P3push}
we compare the time it takes to compute the Tutte polynomials of $P(n,3)$ for 
increasing $n$ using the VORDER-pull (Table \ref{tab:P3pull} and 
VORDER-push (Table \ref{tab:P3push}).
For the first set of timings in each table we apply a full graph isomorphism test for $|V|>15$.
For the second set of timings we identify identical graphs only in the computation tree only.
Column \#calls is the total number of recursive calls made by the algorithm.
Column \#ident counts the number of recursive calls for which the graph is
identical to a graph previously computed in the computation tree.
Column \#isom counts the number of recursive calls for which the graph is not identical but
isomorphic to a graph previously computed in the computation tree.

\begin{table}[htb]
\begin{center}
\begin{tabular}{|l l l| r r r r | r r r |}
\hline
\multicolumn{3}{|c|}{ VORDER-pull}   & \multicolumn{4}{|c|}{with isomorphism test} & \multicolumn{3}{|c|}{no isomorphism test} \\ \hline
$n$ &$|V|$ & $m$ & \#calls & \#ident & \#isom &   time  & \#calls  & \#ident &  time \\ \hline
8   & 16   & 24  &  28641  &  10419  &  0     &  1.21   & 28641    & 10419   & 1.19 \\
9   & 18   & 27  &  30235  &  9818   &  3     &  1.40   & 32693    & 10681   & 1.41 \\
10  & 20   & 30  &  90772  &  31049  &  22    &  4.53   & 240600   & 85017   & 12.16 \\
11  & 22   & 33  &  434402 &  149286 &  244   &  26.63  & 736447   & 259390  & 44.82 \\
12  & 24   & 36  &  471530 &  152284 &  978   &  34.72  & 1217966  & 406976  & 87.89 \\
13  & 26   & 39  & 1668636 & 552034  & 7072   & 177.33  & 5905078  & 2049833 & 730.16 \\
14  & 28   & 42  & 4035615 & 1346519 & 45340  & 798.37  & 17437880 & 6062683 & 2805.78 \\
15  & 30   & 45  & 6330229 & 2016961 & 149699 & 2149.02 &          &         & \\ \hline
\end{tabular}
\end{center}
\caption{Data for $P(n,3)$ for VORDER-pull with the SHARC ordering.} \label{tab:P3pull}
\end{table}
\begin{table}[htb]
\begin{center}
\begin{tabular}{|l l l| r r r r | r r r |}
\hline
\multicolumn{3}{|c|}{ VORDER-push}   & \multicolumn{4}{|c|}{with isomorphism test} & \multicolumn{3}{|c|}{no isomorphism test} \\ \hline
$n$ &$|V|$ & $m$ & \#calls & \#ident & \#isom &   time  & \#calls  & \#ident &  time \\ \hline
 8   & 16   & 24  &  2776   &   703   &    0   &   0.10  &    2776  &   703   &  0.09 \\
 10  & 20   & 30  &  4680   &  1119   &    6   &   0.31  &    6490  &  1634   &  0.23 \\
 12  & 24   & 36  &  7449   &  1828   &   16   &   0.79  &    9552  &  2487   &  0.40 \\
 14  & 28   & 42  & 40142   & 10639   &  192   &   7.66  &   46924  & 12962   &  2.34 \\
 16  & 32   & 48  & 62306   & 16316   &  691   &  26.04  &   77896  & 22103   &  4.41 \\
 18  & 36   & 54  & 88244   & 23154   & 1299   &  55.82  &  112280  & 32545   &  7.30 \\
 20  & 40   & 60  & 115682  & 30503   & 1996   & 105.72  &  148412  & 43676   & 11.30 \\
 22  & 44   & 66  & 143035  & 37734   & 2754   & 181.40  &  184852  & 54925   & 15.68 \\
 24  & 48   & 72  & 170917  & 45204   & 3501   & 289.46  &  221107  & 66114   & 21.70 \\
 26  & 52   & 78  & 198675  & 52641   & 4278   & 445.24  &  257671  & 77437   & 29.50 \\
 28  & 56   & 84  & 226615  & 60085   & 5071   & 674.25  &  294126  & 88700   & 39.07 \\
 30  & 60   & 90  & 254629  & 67585   & 5855   & 975.65  &  330379  & 99888   & 50.28 \\ \hline
\end{tabular}
\end{center}
\caption{Data for $P(n,3)$ for VORDER-push with the SHARC ordering.} \label{tab:P3push}
\end{table}

In comparing the data for $P(n,3)$, it's clear that VORDER-push (Table \ref{tab:P3push}) is much
better than VORDER-pull (Table \ref{tab:P3pull}).  In fact, VORDER-push is polynomial time in $n$.
The reader can see that the number of graphs (column \#calls) is increasing linearly with $n$.
We find the same linear increase for VORDER-push for $P(n,1),$ $P(n,2)$, $P(n,3)$ and $P(n,4)$.
For $P(n,5)$ and $P(n,6)$ the data is not clear.  

The data for $P(n,3)$ also shows that a high percentage of isomorphic graphs in the computation tree are
identical (compare columns \#ident and \#isom).  The data shows that the explicit graph isomorphism test
helps VORDER-pull (Table \ref{tab:P3pull}) but hurts the performance of VORDER-push (Table \ref{tab:P3push}).

In Table \ref{tab:P6} we show data for $P(n,6)$.
The irregularity of the data in Table \ref{tab:P6} is 
partly explained by low girth.  In particular, $P(18,6)$ has 6 triangles.
The girth of $P(n,k)$ is a minimum when $k$ divides $n$ where the girth is $n/k$.
In Table \ref{tab:P14} we fix $n$ to be 14 and vary $k$ to show the dependence
on the girth.

\begin{table}[htb]
\begin{center}
\begin{tabular}{|l l| r r r | r r r |}
\hline
   &       & \multicolumn{3}{|c|}{ VORDER-pull } & \multicolumn{3}{|c|}{ VORDER-push } \\ \hline
$k$& girth & time(s) & \#calls & \#ident  & time(s)   & \#calls & \#ident \\ \hline
13 &   5   &   85.55 &  875232 &    270060&    0.30   &    6884 &  1715  \\
14 &   6   & 1262.37 &  5524084&   1807371&    4.16   &   85103 & 23822  \\
15 &   5   &        &         &           &    6.62   &  124203 & 35033  \\
16 &   7   &        &         &           &   43.69   &  606569 & 177341 \\
17 &   6   &        &         &           &   23.35   &  384107 & 112730 \\
18 &   3   &        &         &           &    3.98   &   65379 & 16181  \\
19 &   6   &        &         &           &   24.55   &  315584 & 87375  \\
20 &   7   &        &         &           &  482.93   & 3647975 & 1081545 \\ \hline
\end{tabular}
\end{center}
\caption{Data for $P(n,6)$ for VORDER-pull and VORDER-push.} \label{tab:P6}
\end{table}

\begin{table}[htb]
\begin{center}
\begin{tabular}{|l l | r r r | r r r |}
\hline
     &    &    \multicolumn{3}{|c|}{ VORDER-pull } & \multicolumn{3}{|c|}{ VORDER-push } \\ \hline
$k$  & girth &  time(s)  &  \#calls  &   deg  &  time(s) &  \#calls &   deg \\ \hline
 1   &   4   &   6.12    &  54040    &  6.48  &    0.16  &     693  &  2.10 \\
 2   &   5   &  209.33   &  1362412  &  5.19  &    0.65  &    4727  &  2.30 \\
 3   &   6   &  806.92   &  4035615  &  4.32  &    3.82  &   40142  &  2.47 \\
 4   &   7   & 2273.75   &  8430139  &  4.61  &    7.71  &   88579  &  2.49 \\
 5   &   6   & 1218.51   &  6208087  &  4.49  &    5.62  &   71717  &  2.50 \\
 6   &   6   &  979.73   &  5524084  &  4.44  &    6.43  &   71054  &  2.47 \\ \hline
\end{tabular}
\end{center}
\caption{Data for $P(14,k).$
Column deg shows the average degree of the first vertex in the computation tree. } \label{tab:P14}
\end{table}

\subsection{The truncated icosahedron graph.}

\begin{center}
\begin{figure}[htb] \centering
\begin{tabular}{cc}
\hspace*{-5mm} \includegraphics[width=0.45\textwidth]{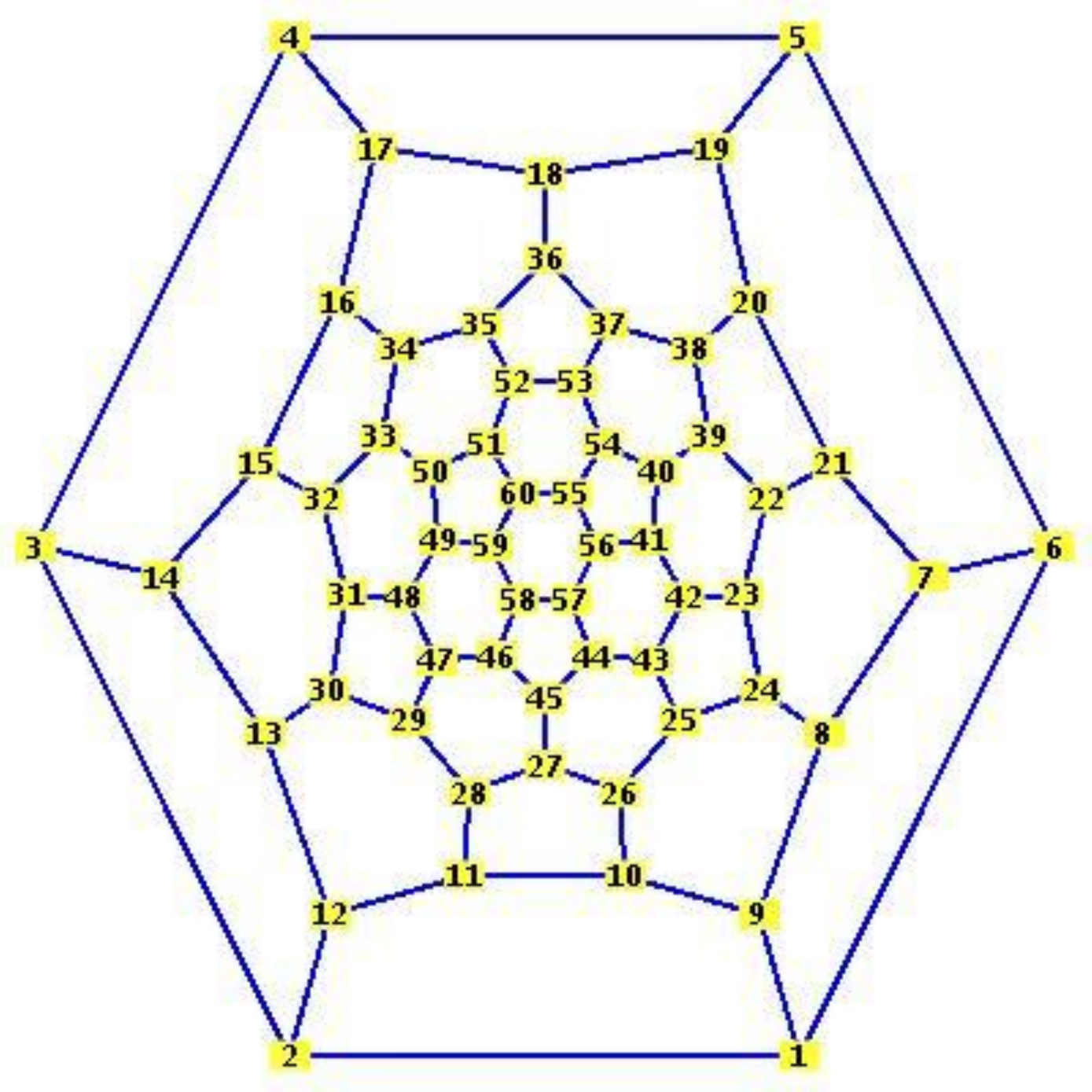}  &
\hspace*{-5mm} \includegraphics[width=0.44\textwidth]{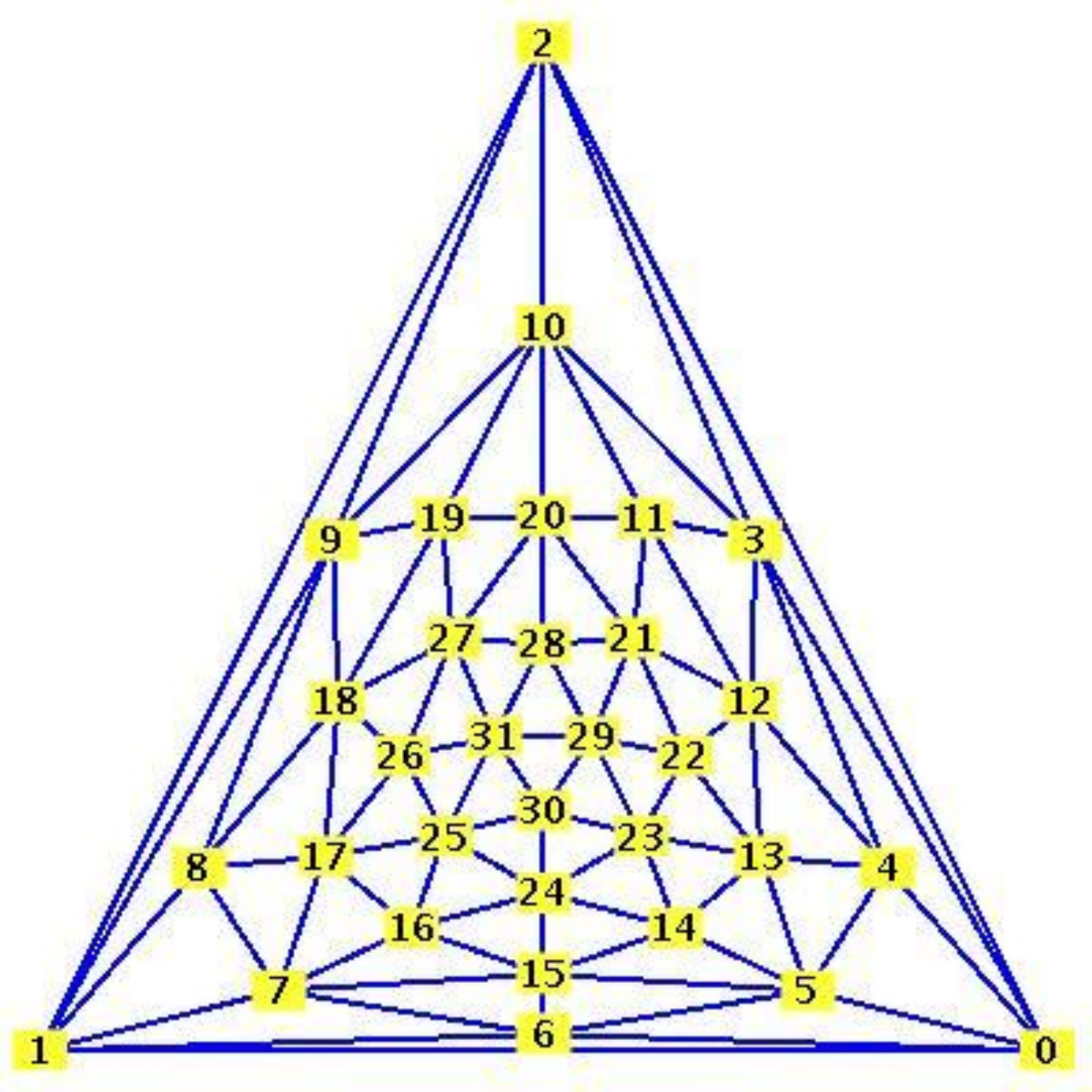}
\end{tabular}
    \caption{The truncated icosahedron graph and its dual.}
    \label{fig:TI}
\end{figure}
\end{center}

The Tutte polynomial of a planar graph $G$ and its dual $G^*$ are related by
$T(G,x,y) = T(G^*,y,x)$.
Shown in Figure \ref{fig:TI} is the truncated icosahedron graph $TI$ and its dual $TI^*$.

In \cite{HPR2011}, Haggard, Pearce and Royle report that they computed the
Tutte polynomial for $TI^*$ in one week on 150 computers.  They used the VORDER-pull heuristic.
Using the VORDER-push heuristic, and the vertex ordering as shown in the figure \ref{fig:TI},
we computed the Tutte polynomial for $TI$ on a single core of a 2.66 Ghz Intel Core i7
desktop in under 4 minutes and 2.8 gigabytes, and for $TI^*$ in under 9 minutes and 8.8 gigabytes.  
Notice that the vertexes of $TI$ (and also $TI^*$) are numbered in concentric cycles.  
This was the ordering that we input the graph from a picture.  Notice that the vertex ordering
is a short arc ordering.  This is why we tried the short arc ordering on other graphs.

\subsection{Dense graphs.}

Up to this point, the data shows that VORDER-push is much better than VORDER-pull.
This, however, is not the case for dense graphs.
In Table \ref{tab:Kn} we give data for the complete graphs $K_n$ on $n$ vertexes.
VORDER-pull is clearly better than VORDER-push.

\begin{table}[htb] \centering
\begin{tabular}{|l l| r r r r | r r r r |}
\hline
     &     & \multicolumn{4}{|c|}{ VORDER-pull }   & \multicolumn{4}{|c|}{ VORDER-push } \\ 
\hline
$n$  & $m$ &  time(s) & \#calls & \#ident& deg     & time   & \#calls  & \#ident &   deg   \\
10   & 45  &    0.08  &   2519  &  1002  &  7.40   &   0.24 &    7448  &   2826  &  4.84   \\
11   & 55  &    0.18  &   5075  &  2024  &  8.27   &   0.64 &   17178  &   6667  &  5.25   \\
12   & 66  &    0.46  &  10191  &  4070  &  9.12   &   1.72 &   38940  &  15372  &  5.70   \\
13   & 78  &    1.07  &  20427  &  8164  & 10.02   &   4.57 &   87070  &  34829  &  6.10   \\
14   & 91  &    2.43  &  40903  & 16354  & 10.90   &  12.64 &  192544  &  77838  &  6.54   \\
15   & 105 &    6.10  &  81859  & 32736  & 11.81   &  39.00 &  421922  & 172047  &  6.95   \\
16   & 120 &   16.36  & 163775  & 65502  & 12.71   & 113.42 &  917540  & 376848  &  7.37   \\ 
17   & 136 &   46.25  & 327611  & 131036 & 13.64   & 273.40 & 1982502  & 819217  &  7.78   \\
18   & 153 &  113.39  & 655287  & 262106 & 14.54   &        &          &         &         \\
\hline
\end{tabular} \caption{Data for $K_n$ for VORDER-pull and VORDER-push} \label{tab:Kn}
\end{table}

\subsection{Monitoring execution for large graphs.}

For large graphs, the user of software for computing Tutte polynomials will need some
way to know how far a large computation has progressed and how much memory has been consumed so that
the user can stop the computation when it becomes obvious that it not going to terminate in a
reasonable time.  When the Tutte polynomial for a graph $G$ of size $n$ 
vertexes in the computation tree is computed for the first time,
we display the additional time it took to compute $T(G)$ since the
time it took to compute the Tutte polynomial for a graph of size $n-1$ for the first time,
and the total space used after $T(G)$ is computed.
In Table \ref{tab:jtoi} and Table \ref{tab:itoj} we show the output of VORDER-push
(VORDER-pull respectively) for the truncated icosahedron $TI$ (for $n>20$).
The reader can see that VORDER-pull will take a very long time.

\begin{center}
\begin{table}[htb] \centering
\begin{tabular}{|rrr|rrr|rrr|rrr|} 
\hline
$n$    &  time(s)   &  space  &  $n$   & time  & space & $n$ & time & space & $n$ & time(s) & space \\ 
\hline
21  &0.06  &0.067gb   &  31  &0.84  &0.100gb  &41  &4.60    &0.258gb &51  &25.32   &1.167gb \\
22  &0.09  &0.090gb   &  32  &0.00  &0.100gb  &42  &5.61    &0.340gb &52  &22.60   &1.443gb \\
23  &0.17  &0.095gb   &  33  &1.32  &0.120gb  &43  &0.00    &0.340gb &53  &0.01    &1.443gb \\
24  &0.00  &0.095gb   &  34  &0.00  &0.120gb  &44  &1.36    &0.360gb &54  &0.20    &1.451gb \\
25  &0.10  &0.095gb   &  35  &0.18  &0.120gb  &45  &5.52    &0.443gb &55  &25.12   &1.786gb \\
26  &0.00  &0.095gb   &  36  &0.00  &0.120gb  &46  &10.98   &0.619gb &56  &12.77   &1.950gb  \\
27  &0.34  &0.095gb   &  37  &1.93  &0.150gb  &47  &0.00    &0.619gb &57  &0.00    &1.950gb \\
28  &0.00  &0.095gb   &  38  &0.00  &0.150gb  &48  &12.69   &0.757gb &58  &6.82    &2.058gb \\
29  &0.55  &0.095gb   &  39  &3.15  &0.193gb  &49  &10.52   &0.880gb &59  &43.38   &2.679gb \\
30  &0.00  &0.095gb   &  40  &0.00  &0.193gb  &50  &0.01    &0.880gb &60  &8.13    &2.761gb \\
\hline
\end{tabular}
    \caption{Trace of time and space for the truncated icosahedron using VORDER-push.
             Total time 204.58 seconds.} \label{tab:jtoi}
\end{table}
\end{center}

\begin{center}
\begin{table}[htb] \centering
\begin{tabular}{|rrr|rrr|rrr|rrr|} 
\hline
$n$    &  time(s)   &  space  &  $n$   & time  & space & $n$ & time & space & $n$ & time(s) & space \\ 
\hline
21 &   0.08s  & 0.085gb &27 &   1.04s  & 0.099gb &33 &   7.44s  & 0.241gb & 39 & 205.00s  & 1.532gb \\
22 &   0.00s  & 0.085gb &28 &   0.00s  & 0.099gb &34 &   0.00s  & 0.241gb & 40 &   0.10s  & 1.532gb \\
23 &   0.36s  & 0.095gb &29 &   1.88s  & 0.115gb &35 &  14.77s  & 0.340gb & 41 & 399.84s  & 3.115gb \\
24 &   0.00s  & 0.095gb &30 &   0.00s  & 0.115gb &36 &   0.20s  & 0.341gb & 42 &   0.01s  & 3.115gb \\
25 &   0.73s  & 0.095gb &31 &   5.22s  & 0.160gb &37 &   0.00s  & 0.341gb & 43 & 758.37s  & 6.205gb   \\
26 &   0.00s  & 0.095gb &32 &   0.00s  & 0.160gb &38 &  59.27s  & 0.661gb & 44 & $>$1500s & $>$14gb \\
\hline
\end{tabular}
    \caption{Trace of time (in seconds) and space for the truncated icosahedron
             using VORDER-pull.} \label{tab:itoj}
\end{table}
\end{center}

\section{Conclusion}

We have presented a new edge selection heuristic that we call VORDER-push for computing
the Tutte polynomial of a graph using the edge deletion and contraction algorithm.
We find that for sparse graphs, VORDER-push outperforms VORDER-pull and the other
heuristics considered by Haggard, Pearce and Royle in \cite{HPR2010}
by several orders of magnitude and which significantly increases the range of graphs that
can be computed for what is an NP-hard problem.  For some graphs, including grid graphs and the
Petersen graphs $P(n,k)$ for $1\le k \le 4$, our new heuristic automatically
finds polynomial time constructions for the Tutte polynomial.
At this point we only have a partial understanding of why and when VORDER-push is so effective.
Graphs with large girth appear to be more difficult.

We are integrating our new heuristic into the {\tt TuttePolynomial} command
in Maple's GraphTheory package.  This should become available in Maple 17.
The overall improvement is huge.  For example, for the dodecahedron graph,
a cubic graph with 20 vertices and 30 edges, the time to compute the
Tutte polynomial improves from 162.093 seconds in Maple 16 to 0.219 seconds.
The GraphTheory package \cite{farr,moh} has been under development since 2004.
We have also installed a command for computing the reliablity polynomial $R_p(G)$ in the package.
Our improvement for computing $T(G,x,y)$ will automatically improve Maple's
performance for computing the chromatic polynomial $P_{\lambda}(G)$ and 
other related polynomials.

\end{document}